\documentclass[twocolumn,showpacs,preprintnumbers,amsmath,amssymb,aps,prx]{revtex4-2}

\usepackage{amsmath}
\usepackage{graphicx}
\usepackage{color}
\usepackage{upgreek}
\usepackage{epstopdf}

\begin{document}

\title{Solitons, chaos, and quantum phenomena: a deterministic approach to the Schrödinger equation
}
\author{Dami\`a Gomila}
\thanks{damia@ifisc.uib-csic.es} 
\affiliation{Instituto de F\'isica Interdisciplinar y Sistemas Complejos, IFISC (CSIC-UIB), Campus Universitat de les Illes Balers, E-07122 Palma de Mallorca, Spain}

\date{\today}

\begin{abstract}
We show that the Schr\"odinger equation describes the ensemble mean dynamics of solitons in a Galilean invariant field theory where we interpret solitons as particles. On a zero background, solitons move classically, following Newton`s second law, however, on a non-zero amplitude chaotic background, their momentum and position fluctuate fulfilling an exact uncertainty relation, which give rise to the emergence of quantum phenomena. The Schrodinger equation for the ensemble of solitons is obtained from this exact uncertainty relation, and the amplitude of the background fluctuations is what corresponds to the value of $\hbar$. We confirm our analytical results running simulations of solitons moving against a potential barrier and comparing the ensemble probabilities with the predictions of the time dependent Schr\"odinger equation, providing a deterministic version of the quantum tunneling effect. We conclude with a discussion of how our theory does not present statistical independence between measurement and experiment outcome. 
\end{abstract}

\maketitle

\section{Introduction}
\label{sec_introduction}
The interpretation of quantum mechanics has been a subject of intense debate since its inception in the mid-1920s, largely due to its probabilistic nature and the concept of wave-particle duality. The Copenhagen interpretation, which is the most widely accepted, posits that quantum mechanics is fundamentally indeterministic, with the wave function amplitude serving as a probability distribution for potential measurement outcomes. In this ontic view, the wave function is the primary entity, and there are no actual particles with defined trajectories until we measure. Despite its widespread acceptance, this statistical description that lacks an underlying physical reality remains controversial \cite{Bush_2020, Hance_2022, Hance_Nat_2022, Hall2024}.

An alternative to the Copenhagen interpretation is the set of hidden variables theories which seek to provide a deterministic framework consistent with the statistical predictions of standard quantum mechanics. Several deterministic interpretations of quantum mechanics have been proposed in the literature \cite{specialissue2021}, including those based on Bohmian mechanics, or quantum pilot-wave theories \cite{Dewdney2023}. These approaches describe particles as moving under the influence of a background field and successfully replicate certain quantum mechanical features. Notably, even macroscopic hydrodynamic pilot-wave systems have been shown to exhibit quantum-like behavior \cite{Bush_2020}. 
However, the separation of the concept of a point particle from the background field in these theories and their explicit nonlocal character introduce certain technical difficulties to make them compatible with special relativity \cite{Drezet_2023, Hatifi2024}. 

A more natural approach could involve solitons, which are solutions to nonlinear field equations and can embody both particle and field characteristics simultaneously. Although solitons have been interpreted as particles within quantum field theory, this has traditionally been done from a standard quantum mechanics perspective \cite{Skyrme1961, Derrick1964, Manton_Sutcliffe_2004, Manton2008, Bowcock_2009, Melo_2016, Lozanov_2023}. The idea of treating solitons as classical particles whose collective dynamics reproduce standard quantum behavior has been relatively unexplored, with only recent studies considering solitons in a pilot-wave context \cite{Drezet_2023}. 

In this paper, we demonstrate that the solitons of a Galilean-invariant nonlinear field can reproduce aspects of the quantum dynamics of microscopic particles in one dimension, as described by the Schrödinger equation (SE), without invoking a pilot wave. Building on an exact soliton uncertainty relation \cite{Hall2002}, we show that the SE accurately captures the ensemble behavior of these solitons, with excellent agreement between averages over deterministic soliton trajectories and the predictions of the time-dependent SE. This provides a local, realistic interpretation of selected quantum features within the scope of our approximations. We further discuss how our framework might remain consistent with violations of Bell inequalities if the assumption of measurement independence is relaxed—an idea that arises naturally here due to the chaotic nature of the background field \cite{Palmer_2009, Hall2010, Hall2011}. Overall, our results contribute to the ongoing discussion surrounding deterministic interpretations of quantum mechanics and offer an alternative perspective that warrants further investigation.

\section{Model and Soliton Solutions}
\label{sec_model}
Starting from a relativistic complex scalar field theory with global $U$(1) symmetry supporting solitons \cite{Bowcock_2009}, in which a complex field $\phi(x,t)$ can be defined by the Lagrangian density
\begin{equation}
    \mathcal{L}=\frac{1}{1-|\phi|^2}\left[ \partial_\mu \phi^*\partial^\mu \phi+ |\phi|^2-|\phi|^4 +V(x) |\phi|^2 \right],
    \label{lagrangian}
\end{equation}
one can obtain from the corresponding equations of motion a Galilean version of the complex sine-Gordon equation (GCSGE) in 1+1 dimensions \cite{Melo_2016}:
\begin{eqnarray}
2m'\partial_t \phi = &-& i (1-|\phi|^2)^2 \phi + i(1-|\phi|^2) \partial_{xx} \phi + \label{csge} \nonumber \\ 
&+& i \phi^* (\partial_x \phi)^2 - i V(x) \phi ,
\end{eqnarray}
where $m'$ is a parameter of the ansatz used in the projection to obtain Galilean invariance from the relativistic Lagrangian density, and $V(x)$ is an external potential acting on $\phi$. In the following we set $m'=1/2$ without loss of generality. Eq.~(\ref{csge}) has a Noether charge associated with the global $U$(1) symmetry that provides robustness to non-topological solitons. With $V(x)=0$, given the Galilean invariance, a soliton moving at an arbitrary velocity can be obtained from a boost of a stationary solution \cite{Bowcock_2009,Melo_2016}. A soliton with amplitude $w'$, moving at velocity $v_0$ is given by:
\begin{eqnarray}
&&\phi_s(x,t,w',v_0)= \nonumber \\
=w' \text{sech}&&{[w'(x-X_s)]}e^{i\frac{v_0}{2}(x-X_s)} e^{i(-1+w'^2+\frac{v_0^2}{4})t},
\label{soliton_sol}
\end{eqnarray}
where $X_s(t)=X_s(0)+v_0t$ is the soliton position. The width of the soliton is $\pi/\sqrt{6w'}$ and its wavelength $4\pi/v_0$. 
Solutions such as (\ref{soliton_sol}) are known as Q-balls in field theory \cite{Bowcock_2009,Melo_2016}, where they are used to describe condensates of a large number of squarks or slepton particles. Here we interpret solitons differently, literally as individual finite-size particles \cite{Manton2008}, in a deterministic way. 

The momentum density of field $\phi$ is 
\begin{equation}
 p(\phi)=\frac{i}{2(1-|\phi|^2)}(\phi \partial_x \phi^* - \phi^* \partial_x \phi).
 \label{momentum}
\end{equation}
Evaluating (\ref{momentum}) for soliton (\ref{soliton_sol}) and integrating over space we obtain that the soliton momentum is given by:
\begin{equation}
 P_s=\int_{-\infty}^\infty p(\phi_s) dx = mv_0,
 \label{momentum_sol}
\end{equation}
where $m=\frac{1}{2}\int_{-\infty}^\infty \frac{|\phi_s|^2}{1-|\phi_s|^2}dx$ is the  mass of the soliton.

In presence of a nonzero external potential gradient, $\partial_x V(x) \neq 0$, form the stress-energy tensor and conservation theorems one obtains 
\begin{equation}
  \partial_t p(\phi_s) = \partial_x J^p(\phi_s)-\frac{|\phi_s|^2}{1-|\phi_s|^2} \partial_x V ,
  \label{eq_cons_mom}
\end{equation}
where $J^p$ is the current associated to the momentum density \cite{Bowcock_2009, Goldstein}. For simplicity, in the rest of this work we consider small particles ($w' \ll 1$) and  gentle potential gradients. Under the latter approximation, the main effect of the potential is changing the soliton velocity, and deformations of the soliton profile can be disregarded. Thus, integrating (\ref{eq_cons_mom}) over space, we obtain Newton's second law:
\begin{equation}
 \dot{P_s} = m\dot{v_0}= -\partial_{X_s} V_{p}(X_s),
 \label{newton_law}
\end{equation}
where 
\begin{equation}
    V_p(X_s)=\int_{-\infty}^\infty \frac{|\phi_s(x-X_s)|^2}{1-|\phi_s(x-X_s)|^2}V(x) dx
    \label{potential_ener}
\end{equation}{}
is the particle's potential energy at position $X_s$. 

\begin{figure}
\includegraphics[width=8cm]{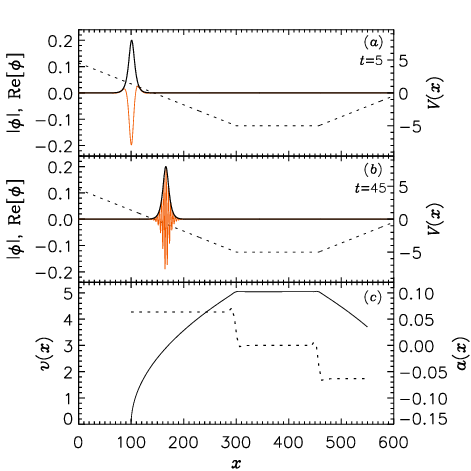}
\caption{Newton's second law for solitons. (a) and (b) show two snapshots of a simulation where a soliton moves on a linear piece-wise potential, $V(x)$, with different slopes. The potential is plotted with a dotted line (right scale). The solid black and orange lines show the amplitude and real part of the field $\phi$ respectively (left scale). Panel (c) shows the velocity, $v_0$ (solid line, left scale), and acceleration, $a=\dot{v}_0$ (dotted line, right scale), of the soliton as a function of position. Note how the envelope of the soliton remains almost unchanged during the acceleration, but the wavenumber of the inner traveling wave increases with the velocity as $k=\frac{v_0}{2}$, as predicted by (\ref{soliton_sol}). Here $w'=0.2$, and the initial position and velocity of the soliton were $X_s=100$ and $v_0=0.1$. Simulations were performed on a domain with periodic boundary conditions of size $L=628.32$ and $4096$ grid points, using a pseudo-spectral method. All quantities are dimensionless. A movie of the time evolution is available in the SI.}
\label{fig:newton_law}
\end{figure}

Newton's second law commonly appears in field theories as an emergent property of solitons \cite{Manton_Sutcliffe_2004}. To illustrate this Newtonian mechanics, we show in Fig. \ref{fig:newton_law} the dynamics of a soliton moving on a linear piece-wise potential with different slopes. We observe how it gets accelerated according to (\ref{newton_law}). As the slope of the potential is gentle, the soliton does barely get deformed by the acceleration, and the only visible effect at naked eye is the increase of the wavenumber $k$ of the field inside the soliton with the velocity as $k=\frac{v_0}{2}$, as predicted by (\ref{soliton_sol}). Higher accelerations can deform the soliton envelope and induce oscillations, or, if the energies involved are high enough, even destroy it, indicating a natural frontier between low and high energy regimes \cite{Manton_Sutcliffe_2004}. In the latter case, the approximations taken to derive Eq.~(\ref{newton_law}) do not hold. Also, the relativistic version of the CSGE \cite{Bowcock_2009, Melo_2016} would be more adequate to describe such regimes.

%%%%%%%%%%%%%%%%%%%%%%%%%%%%%%%%%%%%%%%%%%%%%%%%%%%%%%%%%%%%%%%%%%%%%%%%%%%%%%%%%%%%%%%%%%%%%%%%%%%%%%%%%%%%%%%%%%%%%%%
%%%%%%%%%%%%%%%%%%%%%%%%%%%%%%%%%%%%%%%%%%%%%%%%%%%%%%%%%%%%%%%%%%%%%%%%%%%%%%%%%%%%%%%%%%%%%%%%%%%%%%%%%%%%%%%%%%%%%%%

\section{Dynamics of solitons in a chaotic background}
\label{sec_dyn_fluc}

So far we have proved that in a zero background solitons of field theory (\ref{csge}) follow classical mechanics. Assuming, however, that microscopic particles move in a perfect zero background is not justified {\it a priori}. If the background field is non-zero, the dynamics of solitons changes qualitatively.

%%%%%%%%%%%%%%%%%%%%%%%%%%%%%%%%%%%%%%%%%%%%%%%%%%%%%%%%%%%%%%%%%%%%%%%%%%%%%%%%%%%%%%%%%%%%%%%%%%%%%%%%%%%%%%%%%%%%%%%
\begin{figure}
\includegraphics[width=8cm]{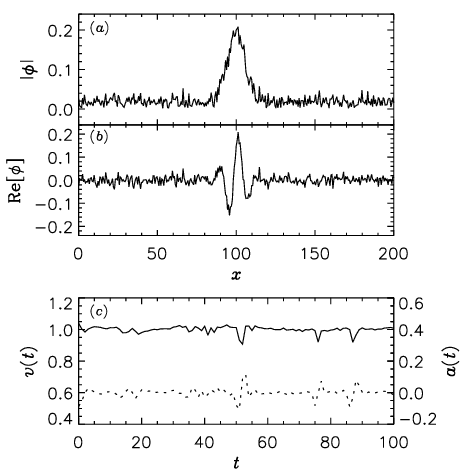}
\caption{Soliton moving in a chaotic background. Panels (a) and (b) show snapshots of the amplitude and real part of the field $\phi$ respectively. Panel (c) shows the velocity, $v$ (solid line), and acceleration, $a$ (dotted line), of the soliton as a function of time. Here $V(x)=0$, $\epsilon=0.02$, and the soliton moves at a constant average velocity $v_0=1$. Other parameters as in Fig.~\ref{fig:newton_law}.}
\label{fig:soliton_fluc}
\end{figure}
%%%%%%%%%%%%%%%%%%%%%%%%%%%%%%%%%%%%%%%%%%%%%%%%%%%%%%%%%%%%%%%%%%%%%%%%%%%%%%%%%%%%%%%%%%%%%%%%%%%%%%%%%%%%%%%%%%%%%%%

Let us consider now how a soliton behaves when placed in a chaotic background. For that, we start a simulation of Eq.~(\ref{csge}) from an initial condition consisting of a soliton in an arbitrary (noisy) background:  
\begin{equation}
 \phi(x,0)=\phi_s(x,0,w',v_0)+ \zeta(x),
 \label{ic}
\end{equation}
where $\zeta(x)$ is a complex white Gaussian noise with correlations $\langle \zeta(x)\zeta(x')\rangle =0$ and $\langle \zeta(x)\zeta^*(x')\rangle =\epsilon^2 \delta(x-x')$ ($\langle  \rangle $ stands for average over noise realizations). This noise mimics an arbitrary initial state of the background. We take here $\epsilon \ll \omega'$. Starting from such initial condition, the background evolves in a spatio-temporally chaotic manner. This is due to the fact that all plane wave solutions of Eq.~(\ref{csge}) are unstable, a condition typically leading to spatio-temporal chaos \cite{Chate1996, Ablowitz1994}. See Appendix \ref{appendix_plane_waves} for the stability analysis of plane waves. The full characterization of the spatio-temporal chaos in Eq.~(\ref{csge}) is left for future studies. Assuming the field can be decomposed at any time in a soliton part and a small amplitude background, 
\begin{equation}
 \phi(x,t) = \phi_s(x,t,w',v_0)+\eta(x,t),
 \label{Psi_perturb}
\end{equation}
from energy conservation, the average squared amplitude of the background is $\langle |\eta(x,t)|^2\rangle =\epsilon^2$, determined by the initial condition.
The soliton remains stable during the evolution and, although slightly deformed by the background field fluctuations, it moves at a constant average velocity $v_0$ (Fig. \ref{fig:soliton_fluc}). Its instantaneous velocity, however, fluctuates in time as a result of the interaction of the soliton with the chaotic background. So does the acceleration around zero. The acceleration fluctuations are an indication of the force exerted by the background on the soliton. 

In the following, we consider that the background field component $\eta(x,t)$ has correlations $\langle \eta(x,t)\eta(x',t')\rangle =0$ and $\langle \eta(x,t)\eta^*(x',t')\rangle =\epsilon^2 \delta(x-x')\delta(t-t')$. This is, strictly speaking, not correct, as $\eta(x,t)$ evolves deterministically according to Eq.~(\ref{csge}). However, if the spatio-temporal chaotic dynamics is fast compared to the soliton dynamics, the decay of correlations over sufficiently short space and time scales can make deterministic fluctuations effectively behave as uncorrelated noise. Such coarse-grained approximations are commonly employed, for example, in modeling turbulent flows  \cite{Frisch1995}, where microscopic determinism generates macroscopic behavior well described by random processes. In this sense, the assumption of delta-correlated fluctuations is not fundamental but rather an effective description valid when the soliton dynamics averages over the underlying chaos. This randomness assumption is key to obtain statistical predictions equivalent to those of the Schrödinger equation. Under this approximation, on average, the force exerted by the background on the soliton has zero expectation value and is uncorrelated with the soliton location. Therefore, it does not perform work on average, hence the constant average velocity. The characteristics of this force recall the concept of quantum force in the Madelung hydrodynamic formulation of the Schrödinger equation \cite{Tsekov_2017,Heifetz_2020,Bush_2020}, and the resulting momentum fluctuations resemble those postulated in stochastic quantum dynamics \cite{Nelson1966}.

\subsection{Solitons' uncertainty principle}
\label{sec_uncertainty}

The velocity fluctuations described above indicate that the momentum of a soliton moving in a spatio-temporally chaotic background fluctuates. 
To evaluate the standard deviation of the momentum fluctuations, $\sigma_P$, we first substitute (\ref{Psi_perturb}) in (\ref{momentum}) using $\phi=\sqrt{\xi_s+\xi}e^{i(\theta_s+\theta)}$, where $\phi_s\equiv \sqrt{\xi_s}e^{i\theta_s}$ and $\xi$ and $\theta$ are the amplitude and phase fluctuations respectively. Note that $\langle \xi\rangle =\epsilon^2$, $\langle \theta\rangle =0$, and $\langle \theta^2\rangle =O(\epsilon^2/w'^2)$. At leading order with $w'$ and $\epsilon$ we obtain that the momentum density fluctuations around the soliton, $\delta p = p(\phi) - p(\phi_s)$, are 
\begin{equation}
 \delta p \approx \xi_s \partial_x \theta.
 \label{momentum_fluc}
\end{equation}
Therefore, the soliton's momentum fluctuations can be evaluated as $\delta P= \int_{-\infty}^\infty \delta p dx$, which leads to (see Appendix \ref{appendix_momentum_fluc})
\begin{equation}
 \sigma^2_P=\langle \delta P^2\rangle =\frac{4}{3} w'^3\epsilon^2.
 \label{mom_uncertainty}
 \end{equation}

Similarly, to evaluate the position fluctuations we define the position $X$ of a soliton in a chaotic background as 
\begin{equation}
\label{position}
 X(t)=\frac{1}{2w'}\int_{X_s-l/2}^{X_s+l/2} x(|\phi(x,t)|^2-\epsilon^2)dx, 
\end{equation}
so that $\langle X(t)\rangle =X_s(t)$. Here we have used that $\int_{-\infty}^\infty |\phi(x,t)|^2 = 2w'$. As the variance of the position fluctuations, $\sigma_X^2$, diverges for $l \rightarrow \infty$, here $l$ must be an appropriate integration domain size, larger than the soliton width but small enough to avoid the divergence (see Appendix \ref{appendix_position_fluc}). Then, we can calculate $\sigma_X^2$ as
\begin{equation}
 \sigma^2_X=\langle \delta X^2\rangle =\frac{\pi^2}{12 \omega'^3}\epsilon^2,
 \label{pos_uncertainty}
\end{equation}
where $\delta X= X-X_s=\frac{1}{2w'}\int_{X_s-l/2}^{X_s+l/2} x(\xi-\epsilon^2)dx$.
Thus, one obtains an exact uncertainty principle for the solitons of (\ref{csge}): 
\begin{equation}
\label{sol_uncertainty}
 \sigma_X \sigma_P=\frac{1}{3}\pi \epsilon^2.
\end{equation}

\subsection{Solitons' Schr\"odinger equation}

Alternatively to the standard quantum mechanics interpretation, the Sch\"odinger equation is equivalent to the Hamilton-Jacobi formulation of the mechanics of an ensemble of classical particles subject to {\it random} momentum fluctuations. But, in order to find an exact correspondence between classical and quantum mechanics, the strength of these momentum fluctuations must be precisely determined by the uncertainty in position of the particles, in such a way that they fulfill the exact uncertainty principle \cite{Hall2002}:
\begin{equation}
\label{exact_uncertainty}
 \sigma_X\sigma_P= \frac{\hbar}{2}.
\end{equation}

Thus, hypothetical classical particles of mass $m$ whose momentum and position fluctuations fulfil (\ref{exact_uncertainty}) follow, as an ensemble, the Schr\"odinger equation
\begin{equation}
 \partial_t \psi = i \frac{\hbar}{2m} \partial_{xx}\psi -i \frac{1}{\hbar} V_{p}(x) \psi.
 \label{se}
\end{equation}
Here, $V_{p}(x)$ is the potential energy of a particle at position $x$ and the wave function is written as $\psi(x,t)=\sqrt{\rho(x,t)}e^{i\frac{S(x,t)}{\hbar}}$, where $\rho(x,t)$ is the probability density function (PDF) of finding a particle at position $x$ at time $t$ and $S(x,t)$ is the momentum potential, related to the local mean velocity field $u(x,t)$ that describes the motion of the particles by 
\begin{equation}
 u(x,t)=\frac{1}{m}\partial_x S.
\end{equation}

Note that Eq.~(\ref{se}) is obtained from Eq.~(\ref{exact_uncertainty}) under very general assumptions regarding randomness and causality and provides a description of the particle dynamics in terms of ensemble mean equations. In \cite{Hall2002}, momentum fluctuations are introduced as a means of effectively eliminating the notion of trajectories, but no attempt is made to provide a realistic model for such fluctuations. Similarly, the stochastic interpretation of quantum mechanics \cite{Nelson1966} is agnostic about the origin of the momentum fluctuations of microscopic particles. In contrast, the theory presented here shows that such momentum fluctuations arise naturally in solitons moving in a chaotic background. This chaotic background plays a role analogous to that of the thermal fluctuations driving Brownian motion, where the collisions of fluid molecules impart random kicks to the Brownian particle. In this analogy, the soliton corresponds to the Brownian particle, while the background field acts like the thermal bath, transferring momentum in a deterministic yet unpredictable manner.

Therefore, comparing Eq.~(\ref{sol_uncertainty}) with Eq.~(\ref{exact_uncertainty}), one can conclude that the ensemble dynamics of the solitons governed by Eq.~(\ref{csge}) follow the Schr\"odinger equation with 
\begin{equation}
 \hbar=\frac{2}{3}\pi \epsilon^2.
\end{equation}
The soliton's Schrödinger equation then reads
\begin{equation}
 \partial_t \psi = i \frac{\pi \epsilon^2}{3 w'} \partial_{xx}\psi -i \frac{3}{2\pi \epsilon^2} V_p(x)\psi.
 \label{sol_se}
\end{equation}
Here we have used that, for $w'\ll 1$, the soliton mass can be approximated by $m=w'$. Within this approximation the potential energy is $V_p(x)=\int_{-\infty}^\infty |\phi_s(x-x')|^2 V(x') dx'$. 

To compute the wave function of the soliton ensemble from soliton trajectories we repeat a simulation starting from initial condition (\ref{ic})
a large number of times, each with a different realization of the initial noise $\zeta(x)$. We compute the PDF, $\rho(x,t)$, of finding a soliton at position $x$ at time $t$ as 
\begin{equation}
  \rho(x,t)= \langle \delta(x-X(t))\rangle ,
  \label{probdensity}
\end{equation}
and the local mean velocity, $u(x,t)$, as \cite{Tsekov_2017}
\begin{equation}
 u(x,t)= \frac{\langle v\delta(x-X(t))\rangle }{\rho},
 \label{meanvel}
\end{equation}
where, from (\ref{momentum}) and (\ref{momentum_sol}), we evaluate the instantaneous soliton velocity $v$ as
\begin{eqnarray}
\label{inst_vel}
v=\frac{-1}{w'}\int_{X_s-l/2}^{X_s+l/2} \partial_x (\theta_s+\theta) (|\phi(x,t)|^2-\epsilon^2)dx 
\end{eqnarray}
Then the wave function of an ensemble of solitons is given by $\psi=\sqrt{\rho}e^{i\frac{3S}{2\pi \epsilon^2}}$ with $\partial_x S= w' u$. Note that the de Broglie wavelength of a soliton moving with a momentum $P_s=w'v_0$ reads in our case $\lambda=h/P_s$ with $h=\frac{4}{3}\pi^2 \epsilon^2$, as expected.

\section{Soliton ``quantum'' tunneling effect}
\label{sec_tunneling}

\begin{figure}
\includegraphics[width=8cm]{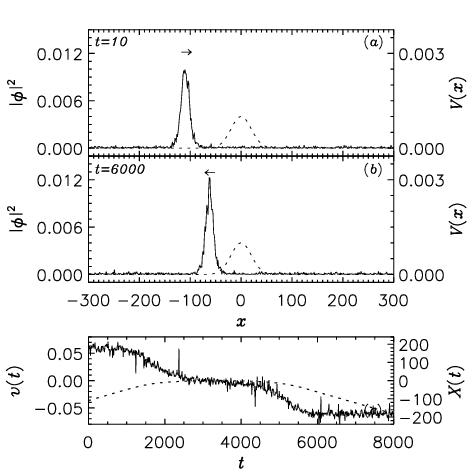}
\caption{Soliton bouncing off a potential barrier. 
Panels (a) and (b) show snapshots of the field amplitude squared, $|\phi|^2$, before and after bouncing off. The potential $V(x)$ (dotted line) is given by (\ref{potential}) with $V_0=0.001$ and $\sigma_V=20$. Panel (c) shows the velocity $v(t)$ (solid line), and position $X(t)$ (dotted line) of the soliton as a function of time.  Solitons are initialized at position $X_s(0)=-110$ with a velocity $v_0=0.0604$, such that their kinetic energy is just below the potential energy of the soliton on top of the barrier. For the chosen parameters, the critical velocity to overcome the barrier is $v_c \approx 2\sqrt{V_0}= 0.063$. Here $w'=0.1$, $\epsilon=0.01$, $L=1256.636$, and the simulation is done with 1024 grid points. A movie of the time evolution is available in the SI.}
\label{fig:tunneling1}
\end{figure}

\begin{figure}
\includegraphics[width=8cm]{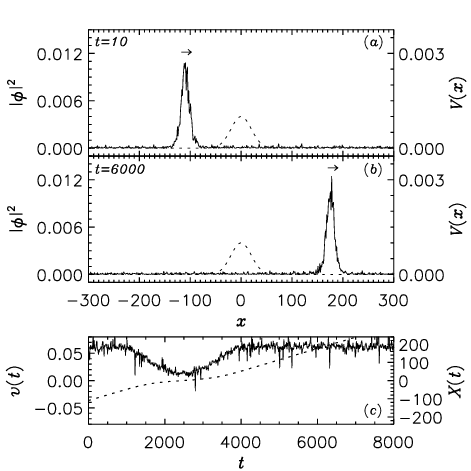}
\caption{Same as in Fig. \ref{fig:tunneling1} for a soliton tunneling through the barrier. A movie of the time evolution is available in the SI.}
\label{fig:tunneling2}
\end{figure}

To confirm our analytical results, we have launched 4500 solitons against a bell shaped potential barrier performing numerical simulations of Eq.~(\ref{csge}) starting from initial condition (\ref{ic}) with a different realization of the initial noise $\zeta(x)$ each time. The potential barrier we have used is
\begin{equation}
 V(x)=V_0 e^{-\frac{x^2}{2\sigma^2_V}},
 \label{potential}
\end{equation}
where $V_0$ and $\sigma_V$ are the height and width of the barrier respectively. We chose an initial velocity of the solitons such that their average kinetic energy, $E_K=\frac{1}{2}mv_0^2$, is lower but close to the potential energy of the soliton on top of the barrier. Then, more than half of the solitons bounce off the barrier but a fraction tunnel through. Figs. \ref{fig:tunneling1} and \ref{fig:tunneling2} show examples of trajectories of a reflected and a transmitted soliton respectively. Whether a soliton bounces off or passes through depends on the specific realization of the spatiotemporally chaotic background fluctuations. 

%%%%%%%%%%%%%%%%%%%%%%%%%%%%%%%%%%%%%%%%%%%%%%%%%%%%%%%%%%%%%%%%%%%%%%%%%%%%%%%%%%%%%%%%%%%%%%%%%%%%%%%%%%%%%%%%%%%%%%%
%%%%%%%%%%%%%%%%%%%%%%%%%%%%%%%%%%%%%%%%%%
%%%%%%%%%%%%%%%%%%%%%%%%%%%%%%%%%%%%%%%%%%
%%%%%%%%%%%%%%%%%%%%%%%%%%%%%%%%%%%%%%%%%%%%%%%%%%%%%%%%%%%%%%%%%%%%%%%%%%%%%%%%%%%%%%%%%%%%%%%%%%%%%%%%%%%%%%%%%%%%%%%

\begin{figure}
\includegraphics[width=8cm]{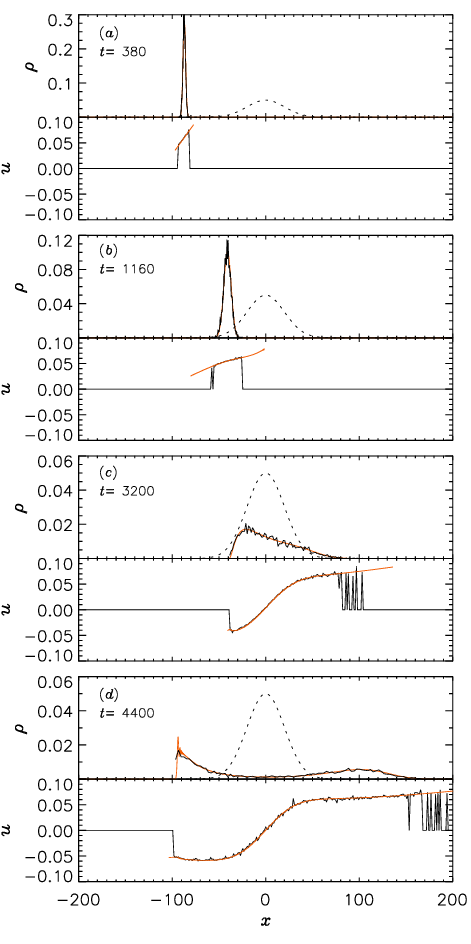}
\caption{Quantum tunneling: comparison of the ensemble mean dynamics obtained integrating Eq.~(\ref{csge}) and averaging for 4500 soliton trajectories (solid black lines), and using the SE (\ref{sol_se}) (orange lines). We plot the probability density function $\rho$ and the local mean velocity $u$ at different times during the evolution. The dotted line shows the potential barrier (\ref{potential}) in an arbitrary scale as reference. Parameters as in Fig. \ref{fig:tunneling1}. Simulations of the SE are done in a domain of size $L=628.32$ with 16384 grid points. 
A movie of the time evolution is available in the SI.}
\label{fig:tunneling3}
\end{figure}

Using the set of all 4500 simulations we compute the PDF, $\rho$, and local mean velocity, $u$, using (\ref{probdensity}) and (\ref{meanvel}) respectively. The obtained ensemble mean dynamics is shown in Fig. \ref{fig:tunneling3} with black solid lines. Simulations of Eq. (\ref{csge}) start at $t=0$ with a soliton located at $X_s=-110$. Due to the noise in the initial condition, the initial PDF is given by $\rho(x,t=0)=\frac{1}{\sigma_X\sqrt{2\pi}} e^{-\frac{(x-X_s)^2}{2\sigma_X^2}}$, with the position uncertainty $\sigma_X$ given by (\ref{pos_uncertainty}). The initial average velocity of solitons is $v_0=0.0604$. The momentum uncertainty is given by (\ref{mom_uncertainty}).

First, we observe that solitons move towards the barrier at the given initial average velocity. As they propagate, the PDF spreads due to momentum uncertainty—faster solitons advance while slower ones lag behind. This effect is reflected in the tilted local mean velocity (Figs. \ref{fig:tunneling3}a and b). The mean velocity at the PDF maximum is $u=v_0$. For practical reasons related to plotting, we set $u=0$ at the locations where we do not find any particle in the simulations. 

When the solitons reach the potential barrier, they either  bounce off or overcome it. Due to the slowing down of solitons in between the classical returning point and the potential maximum, there is a large dispersion in the amount of time they spend in that region. For example, compare the time solitons spend around $x=0$ in Figs. \ref{fig:tunneling1} and \ref{fig:tunneling2}. These dynamics could shed light on the time spent by tunneling particles within the barrier region \cite{Ramos_2020}. 
As a result, $\rho$ spreads considerably in this region and eventually splits in two distinct lumps (see panels (c-d) in Fig. \ref{fig:tunneling3}): one corresponding to solitons that have bounced off, and another to those that have tunneled through. These two lumps move with opposite velocities as indicated by $u$, which goes through zero in between the returning point and the maximum of the potential. The transmission coefficient is given by the the area of the transmitted part of the wavefunction, which for the parameters considered in Fig. \ref{fig:tunneling3} is 0.495.

%%%%%%%%%%%%%%%%%%%%%%%%%%%%%%%%%%%%%%%%%%
%%%%%%%%%%%%%%%%%%%%%%%%%%%%%%%%%%%%%%%%%%%%%%%%%%%%%%%

In orange we plot the results of integrating the SE (\ref{sol_se}) starting from the corresponding initial condition $\psi(x,0)=\sqrt{\rho(x,0)}e^{i\frac{3w'}{2\pi\epsilon^2}v_0x}$. In this case we plot $u$ only where $|\psi|$ is large enough, otherwise the error in computing the phase $S$ is very large. We observe an excellent agreement between the ensemble dynamics and the results from the SE, being the small discrepancies attributable to the limited statistics. The transmission coefficient obtained from the SE is 0.492. We recall that the derivation of the SE is very general, provided the exact uncertainty principle (\ref{sol_uncertainty}) holds \cite{Hall2002}. Differences could arise for larger values of $w'$, $v_0$, or $\epsilon$, as higher order corrections beyond the leading order considered to derive (\ref{sol_uncertainty}) should be taken into account. These nonlinear effects could manifest as deviations from the SE.  

\section{Statistical dependence between measurement and experiment outcome}
\label{superdeterminism}

We have proposed a deterministic, local theory that reproduces certain features of quantum mechanics as described by the Schrödinger equation in one dimension. Under specific approximations, the SE can be derived from this framework. While Bell's theorem precludes local hidden-variable models from reproducing all quantum predictions under standard assumptions, we point out that the assumption of statistical independence between measurement settings and hidden variables may not hold in our model due to the influence of the chaotic background. This raises the possibility that Bell inequality violations might still be consistent with the framework, as suggested in earlier studies \cite{Palmer_2009, Hall2010, Hall2011}.

To illustrate how correlations between measurement and experiment outcome arise in this theory, consider the simulation in Fig.~\ref{fig:tunneling1}, where a soliton collides with a potential barrier and is reflected. The outcome—whether the soliton bounces off or crosses the barrier—is determined by the fluctuations in the background field (i.e., the hidden variables). For example, in Fig.~\ref{fig:tunneling2}, a different realization of background fluctuations causes the soliton to pass through the barrier instead of being reflected.

To demonstrate statistical dependence between measurement and outcome, we conducted paired simulations with identical initial conditions. However, in one of the simulations, we introduced an additional static soliton—representing the probe of a measuring device—positioned behind the moving soliton to detect whether it would bounce off or pass through the barrier. While in some cases the probe did not affect the outcome, in other cases, it did.  

\begin{figure}
\includegraphics[width=8cm]{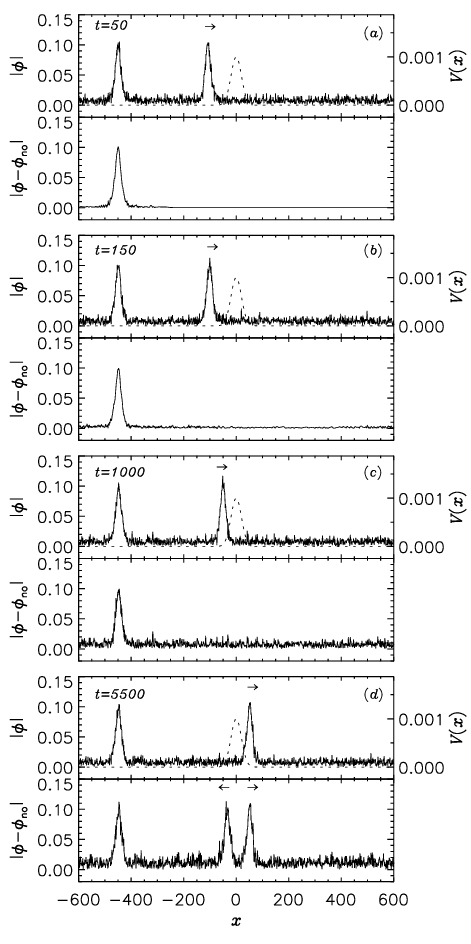}
\caption{Statistical dependence between the measurement performed and the measurement outcome. Panels (a-d) show the time evolution of a soliton starting from exactly the same initial condition than in Fig.~\ref{fig:tunneling1} except for having introduced an extra static soliton at $x= -450$ representing a measuring device. The bottom figure of each panel shows the difference between the field $\phi$ in this simulation and the field in Fig.~\ref{fig:tunneling1} ($\phi_{no}$) with no measuring device. We observe how the simple fact of including a measuring device might change the outcome of the experiment. A movie of the time evolution is available in the SI.}
\label{fig:stat_ind}
\end{figure}

This is exemplified in Fig.~\ref{fig:stat_ind}, where we repeated the simulation shown in Fig. \ref{fig:tunneling1} but included a static "measuring" soliton at $x=-450$. From Fig. \ref{fig:tunneling1}, we know that without measurement, the soliton bounces off the barrier. However, in this case, just the presence of the additional soliton changes the outcome, and the moving soliton passes through the barrier. We computed the difference between the field $\phi$ in this simulation and that of the original simulation without the measuring device, $\phi_{no}$. Over time, we observed that the difference grows, driven by the spatio-temporal chaotic nature of the background fluctuations. Initially, the difference originates from the measuring device, then propagates throughout the system, leading to completely different fluctuating backgrounds over longer time periods.

The simple presence of the measuring device thus modifies the background fluctuations, which in turn alters the experiment’s outcome. This demonstrates the mechanism by which correlations arise between the measurements performed and the hidden variables determining the result.

%%%%%%%%%%%%%%%%%%%%%%%%%%%%%%%%%%%%%%%%%%
%%%%%%%%%%%%%%%%%%%%%%%%%%%%%%%%%%%%%%%%%%%%%%%%%%%%%%%

The divergence between the two chaotic backgrounds arises at the measuring soliton and propagates throughout the system at a finite velocity. This effect is often used to justify loophole-free violations of Bell inequalities \cite{Aspect2015, Storz2023}, as measurements are performed at a sufficient distance to prevent any subluminal signal from influencing the outcome. Thus, it is argued that the measurement and the outcome are independent. However, in certain frameworks \cite{Hooft2017,Hance_Nat_2022,Hossenfelder2020, Bush2024}, as in the one proposed here, this reasoning does not hold. In any real system, measurements cannot be considered independent of past events. Any measurement requires prior actions, allowing the system’s chaotic fluctuations to evolve before the outcome is determined.

An alternative perspective on this issue is provided by the {\it invariant set postulate} \cite{Palmer_2009,Palmer2022,Hossenfelder2020}. Due to the spatio-temporal chaotic nature of the background field fluctuations, the system as a whole moves along a chaotic invariant set. The question of statistical independence then reduces to whether it is possible to find an alternative trajectory in the chaotic invariant set in which, despite the presence of a measuring device, all other degrees of freedom—including the background fluctuations—remain identical to those in the absence of the measurement apparatus at the exact moment when the outcome is determined. In such a chaotic system, this trajectory may not exist, thereby precluding any counterfactual measurement.

The presence of correlations between measurement settings and hidden variables within the invariant set may undermine the assumption of statistical independence, potentially allowing for violations of Bell inequalities. However, testing this hypothesis would require modeling entangled soliton pairs and simulating their joint dynamics, which lies beyond the present scope. Further theoretical and numerical investigations will be necessary to assess the viability of this possibility.

\section{Conclusions}

We have demonstrated that the Schrödinger equation (SE) can be obtained as the ensemble mean dynamics of solitons within a Galilean-invariant field theory. The equation we studied, the GCSGE, represents the simplest known Galilean-invariant field theory that supports solitons. Our findings suggest that similar results could be expected for more complex Galilean or Lorentz-invariant theories, as momentum fluctuations akin to those described in equation (\ref{momentum_fluc}) are likely to arise universally upon linearization, further reinforcing the general applicability of the SE.

Although the GCSGE is a nonlinear partial differential equation, when the energy scales involved are much lower than the soliton's rest energy, the soliton emergent dynamics follows simpler rules: Newton’s second law applies when solitons move through a zero background, while the SE describes the statistical properties of their motion in a nonzero chaotic background. The transition between the two scenarios is straight forward as $\epsilon \rightarrow 0$. Soliton stability is ensured by the conservation of a Noether charge. Furthermore, homogeneous field distributions and plane waves are inherently unstable, resulting in chaotic spatiotemporal background fluctuations. This chaotic background is the source of unpredictability that leads to statistical properties equivalent to those of the SE. 

The agreement between the soliton ensemble dynamics and the derived Schrödinger equation is excellent within the scope of our approximations, providing a local, deterministic framework that reproduces some key aspects of quantum behavior. In this work, we have focused primarily on a deterministic explanation of quantum tunneling, while other phenomena—most notably wavefunction interference, which arguably constitutes the most fundamental manifestation of quantum mechanics—will be explored in future studies. To describe actual fundamental particles, the framework presented here should be extended to higher spatial dimensions. In two and three dimensions, solitons can acquire a topological character—for example, as vortex solitons—and naturally exhibit spin \cite{Manton_Sutcliffe_2004, Pismen1999, Manton2008}. Furthermore, incorporating multiple coupled fields would be necessary to account for additional particle types, interactions, and internal symmetries. While the generalization to such more sophisticated models is technically challenging, the present results establish the groundwork for developing a comprehensive deterministic theory capable of reproducing a broader range of quantum phenomena and potentially connecting with experimentally realizable systems.

While our results are based on a local and deterministic model, they do not necessarily conflict with Bell’s theorem, provided that the assumption of statistical (measurement) independence is relaxed. In our framework, the chaotic background introduces subtle correlations that may violate this assumption. This opens the door to potentially reproducing quantum correlations in a local model, though this remains a debated and unresolved issue. Further work is needed to rigorously examine this possibility.

\acknowledgments
We acknowledge financial support from the Mar\'ia de Maeztu project CEX2021-001164-M funded by the MCIN/AEI/10.13039/501100011033.

\appendix
\section{Linear stability analysis of plane waves}
\label{appendix_plane_waves}

Eq. (\ref{csge}) admits plane wave solutions of the form 
\begin{equation}
 \phi_k(x,t)=w'e^{i(kx-\omega t)},
\end{equation}
with $\omega=1-2w'^2+w'^4+k^2$. 
Substituting $\phi=[w'+\rho(x,t)]e^{i[kx-\omega t + \theta(x,t)]}$ in (\ref{csge}) and keeping only linear terms with $\rho$ and $\theta$ we obtain linearized equations for the evolution of modulus and phase perturbations of a plane wave:
\begin{eqnarray}
\label{linear_eq_plane_waves}
 \partial_t \rho &=& -2k\partial_x \rho -w'(1-w'^2) \partial_{xx} \theta \nonumber \\
 \partial_t \theta &=& 4w'(1-w'^2)\rho + \frac{1-w'^2}{w'} \partial_{xx} \rho -2k \partial_{x} \theta.
\end{eqnarray}
In vector form, the solutions of Eqs. (\ref{linear_eq_plane_waves}) can be written as
\begin{equation}
 \begin{pmatrix} \rho \\ \theta \end{pmatrix}= \vec{v} e^{iqx}e^{\lambda t},
\end{equation}
where $q$ is the wavenumber of the perturbations relative to $k$ and $\lambda$ is the growth rate. Then, stability of plane waves reduces to solving the eigenvalue problem:
\begin{equation}
 \mathcal{J} \vec{v}= \lambda \vec{v} 
\end{equation}
with 
\begin{equation}
 \mathcal{J}=\begin{pmatrix} -i2kq & w'(1-w'^2) q^2 \\ 
 4w'(1-w'^2)-\frac{1-w'^2}{w'}q^2 & -2ikq
 \end{pmatrix}.
\end{equation}
The eigenvalues of the Jacobian $\mathcal{J}$ are given by
\begin{equation}
 \lambda(q)=\pm (1-w'^2)q\sqrt{4w'^2-q^2}-i2kq.
\end{equation}
For a large enough system ($\frac{2\pi}{L} <  2w'$, with $L$ the system size) there is always at least one eigenvalue with positive real part regardless the values of $w'$ and $k$. Therefore all plane waves are unstable in the GCSGE. The wavenumber with maximum growth rate is $q_u=\sqrt{2}w'$.

\section{Momentum uncertainty}
\label{appendix_momentum_fluc}

The instantaneous momentum fluctuations $\delta P(t)$ of soliton (\ref{soliton_sol}) are given by
\begin{equation}
\label{app_momentum_fluc}
 \delta P= \int \delta p dx = \int |\phi_s|^2 \partial_x \theta(x,t) dx.
\end{equation}
Integration limits in all integrals are assumed to be ($-\infty$, $\infty$) unless otherwise indicated. Integrating (\ref{app_momentum_fluc}) by parts, considering that $|\phi_s(\pm \infty)|=0$, we obtain:
\begin{eqnarray}
\label{app_momentum_fluc2}
 &\delta P= - \int \theta \partial_x|\phi_s|^2 dx = & \nonumber \\
 & 2w'^3 \int \theta \text{sech}^2[w'(x-X_s)]\text{tanh}[w'(x-X_s)]dx.&  
\end{eqnarray}

The properties of phase fluctuations $\theta(x,t)$ are directly related to the background field fluctuations $\eta(x,t)$ as follows:
\begin{equation}
 \phi=\phi_s+\eta=\sqrt{\xi_s+\xi}e^{i(\theta_s+\theta)}.
 \label{app_field_fluc}
\end{equation}

Therefore, $\langle \theta\rangle =0$ and, in the core of the soliton, where $|\phi_s|\gg|\eta|$ \footnote{Away from  the soliton's core, where  $|\phi_s|\rightarrow 0$, the phase of $\eta$ must be uniformly distributed over [$-\pi$,$\pi$] and, therefore, $\langle \theta(x,t)\theta(x',t')\rangle =\frac{\pi^2}{3}\delta(x-x')\delta(t-t')$.}, 
\begin{equation}
\label{phase_corr}
 \langle \theta(x,t)\theta(x',t')\rangle =\frac{\epsilon^2 \delta(x-x')\delta(t-t')}{2w'^2\text{sech}^2[w'(x-X_s)]}. 
\end{equation}
Despite (\ref{phase_corr}) diverges for $|x|\rightarrow \infty$, (\ref{app_momentum_fluc2}) remains finite, and we can evaluate the variance of soliton momentum fluctuations as:
\begin{eqnarray}
& \langle \delta P^2\rangle = 4 w'^6 \int \int \langle \theta(x,t)\theta(x',t)\rangle  &\nonumber \\ &\text{sech}^2[w'(x-X_s)]\text{tanh}[w'(x-X_s)] & \nonumber \\
& \text{sech}^2[w'(x'-X_s)]\text{tanh}[w'(x'-X_s)]dxdx' = & \nonumber \\ 
&2 w'^4 \epsilon^2 \int \text{sech}^2[w'(x-X_s)]\text{tanh}^2[w'(x-X_s)] dx =& \nonumber \\
&=\frac{4}{3}  w'^3 \epsilon^2.& 
 \end{eqnarray}

\section{Position uncertainty}
\label{appendix_position_fluc}

%%%%%%%%%%%%%%%%%%%%%%%%%%%%%%%%%%%%%%%%%%
\begin{figure}
\includegraphics[width=8cm]{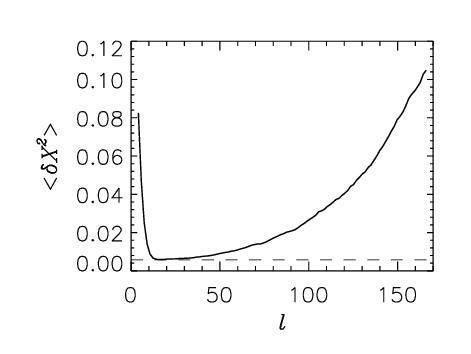}
\caption{Variance of the position fluctuations, $\langle \delta X^2\rangle $, as a function of the integration domain size, $l$. The solid line has been computed numerically evaluating $\langle (X-X_s)^2\rangle $ averaging over 1000 realizations of the noise and calculating the soliton instantaneous position $X$ using (\ref{app_position}). The horizontal dashed line indicates the theoretical value (\ref{app_deltax2}). Here $w'=0.18$ and $\epsilon=0.0064$. For these parameter values, one obtains an accurate enough result for $15 <  l <  30$ approximately.}
\label{fig:uncertl}
\end{figure}
%%%%%%%%%%%%%%%%%%%%%%%%%%%%%%%%%%%%%%%%%%%%%%%%%%%%%%%

We define the position $X$ of a soliton in a fluctuating background as 
\begin{equation}
\label{app_position}
 X(t)=\frac{1}{2w'}\int_{X_s-l/2}^{X_s+l/2} x(|\phi(x,t)|^2-\epsilon^2)dx.
\end{equation}
We recall that $|\phi(x,t)|^2=|\phi_s(x,t)|^2 + \xi(x,t)$. Also, according to (\ref{app_field_fluc}), $\langle \xi\rangle =\epsilon^2$ and $\langle \xi(x,t)\xi(x',t')\rangle =2|\phi_s(x,t)|^2\epsilon^2\delta(x-x')\delta(t-t')+ \langle |\eta(x,t)|^2|\eta(x',t')|^2\rangle $.
%$\langle \xi(x,t)\xi(x',t')\rangle =(2|\phi_s(x,t)|^2\epsilon^2 + 3\epsilon^4)\delta(x-x')\delta(t-t')+2\epsilon^4 (1-\delta(x-x')\delta(t-t'))$. 
Therefore 
\begin{eqnarray}
 \langle  X(t)\rangle =\frac{1}{2w'} \left[ \int_{X_s-l/2}^{X_s+l/2} x|\phi_s(x,t)|^2dx + \right.\nonumber \\
\left. \int_{X_s-l/2}^{X_s+l/2} x(\langle \xi\rangle -\epsilon^2)dx \right]=X_s(t). 
 \label{app_ave_position}
\end{eqnarray}
In what follows, to simplify the notation, we will consider $X_s=0$ by a simple change of variables. Using $\delta X= X-X_s=\frac{1}{2w'}\int_{-l/2}^{l/2} x(\xi-\epsilon^2)dx$, the correlations of the position fluctuations are given by:
\begin{eqnarray}
 \langle \delta X(t)\delta X(t')\rangle = \nonumber \\
 \frac{1}{4w'^2}\int_{-l/2}^{l/2}\int_{-l/2}^{l/2}xx'\langle [\xi(x,t)-\epsilon^2][\xi(x',t')-\epsilon^2]\rangle  dxdx'= \nonumber \\
 \frac{\delta(t-t')}{4w'^2}\left[\epsilon^2\int_{-l/2}^{l/2}2x^2|\phi_s|^2dx +  \langle |\eta|^4\rangle \int_{-l/2}^{l/2} x^2 dx \right] =\nonumber \\
 \frac{\delta(t-t')}{4w'^2}\left[2\epsilon^2w'^2\int_{-l/2}^{l/2}x^2\text{sech}^2(w'x)dx + \frac{\langle |\eta|^4\rangle  l^3}{12} \right] \nonumber \\
 \label{app_pos_fluc1}
\end{eqnarray}
We note that the last term of (\ref{app_pos_fluc1}) diverge when $l\rightarrow \infty$. To avoid this divergence we consider an integration domain size larger than the soliton width  but finite, $\frac{\pi}{\sqrt{6w'}} \ll l$, such that, on the one hand, we can use the following approximation:
\begin{eqnarray}
 &w'^2\int_{-l/2}^{l/2}x^2\text{sech}^2(w'x)dx=\frac{1}{w'} \int_{-lw'/2}^{lw'/2}x'^2\text{sech}^2(x')dx' \approx&  \nonumber \\
 &\frac{1}{w'}\int_{-\infty}^{\infty}x'^2\text{sech}^2(x')dx'= \frac{\pi^2}{6w'},&
\end{eqnarray}
and, on the other hand, we can disregard $\langle  |\eta|^4\rangle  l^3$ with respect to $\frac{4\pi^2\epsilon^2}{w'}$. Assuming $\langle  |\eta|^4\rangle =O(3\epsilon^4)$, we consider $\frac{\pi^3}{(6w')^{\frac{3}{2}}} \ll l^3 \ll \frac{4\pi^2}{3w'\epsilon^2}$, which can be attained if $\epsilon \ll w'$ as assumed in this paper. 
Then, the variance of the position fluctuations is well defined and can be approximated by 
\begin{equation}
 \langle \delta X^2(t)\rangle  = \frac{\epsilon^2 \pi^2}{12w'^3}.
 \label{app_deltax2}
\end{equation}
We have checked numerically that this approximation holds for the parameters used throughout this work (see Fig.~\ref{fig:uncertl}).

%\bibliography{refs}

\begin{thebibliography}{36}%
\makeatletter
\providecommand \@ifxundefined [1]{%
 \@ifx{#1\undefined}
}%
\providecommand \@ifnum [1]{%
 \ifnum #1\expandafter \@firstoftwo
 \else \expandafter \@secondoftwo
 \fi
}%
\providecommand \@ifx [1]{%
 \ifx #1\expandafter \@firstoftwo
 \else \expandafter \@secondoftwo
 \fi
}%
\providecommand \natexlab [1]{#1}%
\providecommand \enquote  [1]{``#1''}%
\providecommand \bibnamefont  [1]{#1}%
\providecommand \bibfnamefont [1]{#1}%
\providecommand \citenamefont [1]{#1}%
\providecommand \href@noop [0]{\@secondoftwo}%
\providecommand \href [0]{\begingroup \@sanitize@url \@href}%
\providecommand \@href[1]{\@@startlink{#1}\@@href}%
\providecommand \@@href[1]{\endgroup#1\@@endlink}%
\providecommand \@sanitize@url [0]{\catcode `\\12\catcode `\$12\catcode
  `\&12\catcode `\#12\catcode `\^12\catcode `\_12\catcode `\%12\relax}%
\providecommand \@@startlink[1]{}%
\providecommand \@@endlink[0]{}%
\providecommand \url  [0]{\begingroup\@sanitize@url \@url }%
\providecommand \@url [1]{\endgroup\@href {#1}{\urlprefix }}%
\providecommand \urlprefix  [0]{URL }%
\providecommand \Eprint [0]{\href }%
\providecommand \doibase [0]{https://doi.org/}%
\providecommand \selectlanguage [0]{\@gobble}%
\providecommand \bibinfo  [0]{\@secondoftwo}%
\providecommand \bibfield  [0]{\@secondoftwo}%
\providecommand \translation [1]{[#1]}%
\providecommand \BibitemOpen [0]{}%
\providecommand \bibitemStop [0]{}%
\providecommand \bibitemNoStop [0]{.\EOS\space}%
\providecommand \EOS [0]{\spacefactor3000\relax}%
\providecommand \BibitemShut  [1]{\csname bibitem#1\endcsname}%
\let\auto@bib@innerbib\@empty
%</preamble>
\bibitem [{\citenamefont {Bush}\ and\ \citenamefont {Oza}(2020)}]{Bush_2020}%
  \BibitemOpen
  \bibfield  {author} {\bibinfo {author} {\bibfnamefont {J.}~\bibnamefont
  {Bush}}\ and\ \bibinfo {author} {\bibfnamefont {A.}~\bibnamefont {Oza}},\
  }\bibfield  {title} {\bibinfo {title} {Hydrodynamic quantum analogs},\
  }\href@noop {} {\bibfield  {journal} {\bibinfo  {journal} {Rep. Prog. Phys.}\
  }\textbf {\bibinfo {volume} {84}},\ \bibinfo {pages} {017001} (\bibinfo
  {year} {2020})}\BibitemShut {NoStop}%
\bibitem [{\citenamefont {Hance}\ \emph {et~al.}(2022)\citenamefont {Hance},
  \citenamefont {Hossenfelder},\ and\ \citenamefont {Palmer}}]{Hance_2022}%
  \BibitemOpen
  \bibfield  {author} {\bibinfo {author} {\bibfnamefont {J.}~\bibnamefont
  {Hance}}, \bibinfo {author} {\bibfnamefont {S.}~\bibnamefont
  {Hossenfelder}},\ and\ \bibinfo {author} {\bibfnamefont {T.}~\bibnamefont
  {Palmer}},\ }\bibfield  {title} {\bibinfo {title} {Supermeasured: Violating
  Bell-statistical independence without violating physical statistical
  independence},\ }\href@noop {} {\bibfield  {journal} {\bibinfo  {journal}
  {Found. Phys.}\ }\textbf {\bibinfo {volume} {52}},\ \bibinfo {pages} {81}
  (\bibinfo {year} {2022})}\BibitemShut {NoStop}%
\bibitem [{\citenamefont {Hance}\ and\ \citenamefont
  {Hossenfelder}(2022)}]{Hance_Nat_2022}%
  \BibitemOpen
  \bibfield  {author} {\bibinfo {author} {\bibfnamefont {J.}~\bibnamefont
  {Hance}}\ and\ \bibinfo {author} {\bibfnamefont {S.}~\bibnamefont
  {Hossenfelder}},\ }\bibfield  {title} {\bibinfo {title} {Bell's theorem
  allows local theories of quantum mechanics},\ }\href@noop {} {\bibfield
  {journal} {\bibinfo  {journal} {Nat. Phys.}\ }\textbf {\bibinfo {volume}
  {18}},\ \bibinfo {pages} {1382} (\bibinfo {year} {2022})}\BibitemShut
  {NoStop}%
\bibitem [{\citenamefont {Hall}(2024)}]{Hall2024}%
  \BibitemOpen
  \bibfield  {author} {\bibinfo {author} {\bibfnamefont {M.~J.~W.}\
  \bibnamefont {Hall}},\ }\bibfield  {title} {\bibinfo {title} {Bell vs. Bell:
  A ding-dong battle over quantum incompleteness},\ }\href
  {https://doi.org/10.3390/foundations4040041} {\bibfield  {journal} {\bibinfo
  {journal} {Foundations}\ }\textbf {\bibinfo {volume} {4}},\ \bibinfo {pages}
  {658} (\bibinfo {year} {2024})}\BibitemShut {NoStop}%
\bibitem [{\citenamefont {Cetto}\ \emph {et~al.}(2021)\citenamefont {Cetto},
  \citenamefont {Casado}, \citenamefont {Hess},\ and\ \citenamefont
  {Vald\'es-Hern\'andez}}]{specialissue2021}%
  \BibitemOpen
  \bibfield  {author} {\bibinfo {author} {\bibfnamefont {A.}~\bibnamefont
  {Cetto}}, \bibinfo {author} {\bibfnamefont {A.}~\bibnamefont {Casado}},
  \bibinfo {author} {\bibfnamefont {K.}~\bibnamefont {Hess}},\ and\ \bibinfo
  {author} {\bibfnamefont {A.}~\bibnamefont {Vald\'es-Hern\'andez}},\
  }\bibfield  {title} {\bibinfo {title} {Editorial: Towards a local realist
  view of the quantum phenomenon},\ }\href@noop {} {\bibfield  {journal}
  {\bibinfo  {journal} {Front. Phys.}\ }\textbf {\bibinfo {volume} {9}},\
  \bibinfo {pages} {651127} (\bibinfo {year} {2021})}\BibitemShut {NoStop}%
\bibitem [{\citenamefont {Dewdney}(2023)}]{Dewdney2023}%
  \BibitemOpen
  \bibfield  {author} {\bibinfo {author} {\bibfnamefont {C.}~\bibnamefont
  {Dewdney}},\ }\bibfield  {title} {\bibinfo {title} {{Rekindling of de
  Broglie–Bohm pilot wave theory in the late twentieth century: A personal
  account}},\ }\href@noop {} {\bibfield  {journal} {\bibinfo  {journal} {Fond.
  Phys.}\ }\textbf {\bibinfo {volume} {53}},\ \bibinfo {pages} {24} (\bibinfo
  {year} {2023})}\BibitemShut {NoStop}%
\bibitem [{\citenamefont {Drezet}(2023)}]{Drezet_2023}%
  \BibitemOpen
  \bibfield  {author} {\bibinfo {author} {\bibfnamefont {A.}~\bibnamefont
  {Drezet}},\ }\bibfield  {title} {\bibinfo {title} {Quantum solitodynamics:
  Non-linear wave mechanics and pilot-wave theory},\ }\href@noop {} {\bibfield
  {journal} {\bibinfo  {journal} {Found. Phys.}\ }\textbf {\bibinfo {volume}
  {53}},\ \bibinfo {pages} {31} (\bibinfo {year} {2023})}\BibitemShut {NoStop}%
\bibitem [{\citenamefont {Hatifi}(2024)}]{Hatifi2024}%
  \BibitemOpen
  \bibfield  {author} {\bibinfo {author} {\bibfnamefont {M.}~\bibnamefont
  {Hatifi}},\ }\bibfield  {title} {\bibinfo {title} {{Relativistic Bohmian
  mechanics revisited: A covariant reformulation for spin-1/2 particles}},\
  }\href {https://doi.org/10.1016/j.physleta.2024.129680} {\bibfield  {journal}
  {\bibinfo  {journal} {Physics Letters A}\ }\textbf {\bibinfo {volume}
  {518}},\ \bibinfo {pages} {129680} (\bibinfo {year} {2024})}\BibitemShut
  {NoStop}%
\bibitem [{\citenamefont {Skyrme}(1961)}]{Skyrme1961}%
  \BibitemOpen
  \bibfield  {author} {\bibinfo {author} {\bibfnamefont {T.}~\bibnamefont
  {Skyrme}},\ }\bibfield  {title} {\bibinfo {title} {A non-linear field
  theory},\ }\href@noop {} {\bibfield  {journal} {\bibinfo  {journal} {Proc. R.
  Soc. Lond. A}\ }\textbf {\bibinfo {volume} {260}},\ \bibinfo {pages}
  {127–138} (\bibinfo {year} {1961})}\BibitemShut {NoStop}%
\bibitem [{\citenamefont {Derrick}(1964)}]{Derrick1964}%
  \BibitemOpen
  \bibfield  {author} {\bibinfo {author} {\bibfnamefont {G.~H.}\ \bibnamefont
  {Derrick}},\ }\bibfield  {title} {\bibinfo {title} {Comments on nonlinear
  wave equations as models for elementary particles},\ }\href
  {https://doi.org/10.1063/1.1704233} {\bibfield  {journal} {\bibinfo
  {journal} {Journal of Mathematical Physics}\ }\textbf {\bibinfo {volume}
  {5}},\ \bibinfo {pages} {1252} (\bibinfo {year} {1964})}
  \BibitemShut {NoStop}%
\bibitem [{\citenamefont {Manton}\ and\ \citenamefont
  {Sutcliffe}(2004)}]{Manton_Sutcliffe_2004}%
  \BibitemOpen
  \bibfield  {author} {\bibinfo {author} {\bibfnamefont {N.}~\bibnamefont
  {Manton}}\ and\ \bibinfo {author} {\bibfnamefont {P.}~\bibnamefont
  {Sutcliffe}},\ }\href@noop {} {\emph {\bibinfo {title} {Topological
  Solitons}}},\ Cambridge Monographs on Mathematical Physics\ (\bibinfo
  {publisher} {Cambridge University Press},\ \bibinfo {year}
  {2004})\BibitemShut {NoStop}%
\bibitem [{\citenamefont {Manton}(2008)}]{Manton2008}%
  \BibitemOpen
  \bibfield  {author} {\bibinfo {author} {\bibfnamefont {N.~S.}\ \bibnamefont
  {Manton}},\ }\bibfield  {title} {\bibinfo {title} {Solitons as elementary
  particles: a paradigm scrutinized},\ }\href@noop {} {\bibfield  {journal}
  {\bibinfo  {journal} {Nonlinearity}\ }\textbf {\bibinfo {volume} {21}},\
  \bibinfo {pages} {T221} (\bibinfo {year} {2008})}\BibitemShut {NoStop}%
\bibitem [{\citenamefont {Bowcock}\ \emph {et~al.}(2009)\citenamefont
  {Bowcock}, \citenamefont {Foster},\ and\ \citenamefont
  {Sutcliffe}}]{Bowcock_2009}%
  \BibitemOpen
  \bibfield  {author} {\bibinfo {author} {\bibfnamefont {P.}~\bibnamefont
  {Bowcock}}, \bibinfo {author} {\bibfnamefont {D.}~\bibnamefont {Foster}},\
  and\ \bibinfo {author} {\bibfnamefont {P.}~\bibnamefont {Sutcliffe}},\
  }\bibfield  {title} {\bibinfo {title} {Q-balls, integrability and duality},\
  }\href@noop {} {\bibfield  {journal} {\bibinfo  {journal} {J. Phys. A: Math.
  Theor.}\ }\textbf {\bibinfo {volume} {42}},\ \bibinfo {pages} {085403}
  (\bibinfo {year} {2009})}\BibitemShut {NoStop}%
\bibitem [{\citenamefont {de~Melo}\ \emph {et~al.}(2016)\citenamefont
  {de~Melo}, \citenamefont {de~Montigny}, \citenamefont {Pinfold},\ and\
  \citenamefont {Tuszynski}}]{Melo_2016}%
  \BibitemOpen
  \bibfield  {author} {\bibinfo {author} {\bibfnamefont {G.}~\bibnamefont
  {de~Melo}}, \bibinfo {author} {\bibfnamefont {M.}~\bibnamefont
  {de~Montigny}}, \bibinfo {author} {\bibfnamefont {J.}~\bibnamefont
  {Pinfold}},\ and\ \bibinfo {author} {\bibfnamefont {J.}~\bibnamefont
  {Tuszynski}},\ }\bibfield  {title} {\bibinfo {title} {{Symmetries and soliton
  solutions of the Galilean complex sine-Gordon equation}},\ }\href@noop {}
  {\bibfield  {journal} {\bibinfo  {journal} {Phys. Lett. A}\ }\textbf
  {\bibinfo {volume} {380}},\ \bibinfo {pages} {1223} (\bibinfo {year}
  {2016})}\BibitemShut {NoStop}%
\bibitem [{\citenamefont {Lozanov}\ \emph {et~al.}(2025)\citenamefont
  {Lozanov}, \citenamefont {Sasaki},\ and\ \citenamefont
  {Takhistov}}]{Lozanov_2023}%
  \BibitemOpen
  \bibfield  {author} {\bibinfo {author} {\bibfnamefont {K.~D.}\ \bibnamefont
  {Lozanov}}, \bibinfo {author} {\bibfnamefont {M.}~\bibnamefont {Sasaki}},\
  and\ \bibinfo {author} {\bibfnamefont {V.}~\bibnamefont {Takhistov}},\
  }\bibfield  {title} {\bibinfo {title} {Universal gravitational wave
  signatures of cosmological solitons},\ }\href
  {https://doi.org/10.1088/1475-7516/2025/01/094} {\bibfield  {journal}
  {\bibinfo  {journal} {Journal of Cosmology and Astroparticle Physics}\
  }\textbf {\bibinfo {volume} {2025}}\bibinfo  {number} { (01)},\ \bibinfo
  {pages} {094}}\BibitemShut {NoStop}%
\bibitem [{\citenamefont {Hall}\ and\ \citenamefont
  {Reginatto}(2002)}]{Hall2002}%
  \BibitemOpen
\bibfield  {number} {  }\bibfield  {author} {\bibinfo {author} {\bibfnamefont
  {M.~J.}\ \bibnamefont {Hall}}\ and\ \bibinfo {author} {\bibfnamefont
  {M.}~\bibnamefont {Reginatto}},\ }\bibfield  {title} {\bibinfo {title}
  {{Schr\"odinger equation from an exact uncertainty principle}},\ }\href@noop {}
  {\bibfield  {journal} {\bibinfo  {journal} {J. Phys. A: Math. Gen.}\ }\textbf
  {\bibinfo {volume} {35}},\ \bibinfo {pages} {3289} (\bibinfo {year}
  {2002})}\BibitemShut {NoStop}%
\bibitem [{\citenamefont {Palmer}(2009)}]{Palmer_2009}%
  \BibitemOpen
  \bibfield  {author} {\bibinfo {author} {\bibfnamefont {T.}~\bibnamefont
  {Palmer}},\ }\bibfield  {title} {\bibinfo {title} {The invariant set
  postulate: a new geometric framework for the foundations of quantum theory
  and the role played by gravity},\ }\href@noop {} {\bibfield  {journal}
  {\bibinfo  {journal} {Proc. R. Soc. A}\ }\textbf {\bibinfo {volume} {465}},\
  \bibinfo {pages} {3165} (\bibinfo {year} {2009})}\BibitemShut {NoStop}%
\bibitem [{\citenamefont {Hall}(2010)}]{Hall2010}%
  \BibitemOpen
  \bibfield  {author} {\bibinfo {author} {\bibfnamefont {M.~J.~W.}\
  \bibnamefont {Hall}},\ }\bibfield  {title} {\bibinfo {title} {Local
  deterministic model of singlet state correlations based on relaxing
  measurement independence},\ }\href
  {https://doi.org/10.1103/PhysRevLett.105.250404} {\bibfield  {journal}
  {\bibinfo  {journal} {Phys. Rev. Lett.}\ }\textbf {\bibinfo {volume} {105}},\
  \bibinfo {pages} {250404} (\bibinfo {year} {2010})}\BibitemShut {NoStop}%
\bibitem [{\citenamefont {Hall}(2011)}]{Hall2011}%
  \BibitemOpen
  \bibfield  {author} {\bibinfo {author} {\bibfnamefont {M.~J.~W.}\
  \bibnamefont {Hall}},\ }\bibfield  {title} {\bibinfo {title} {Relaxed Bell
  inequalities and Kochen-Specker theorems},\ }\href
  {https://doi.org/10.1103/PhysRevA.84.022102} {\bibfield  {journal} {\bibinfo
  {journal} {Phys. Rev. A}\ }\textbf {\bibinfo {volume} {84}},\ \bibinfo
  {pages} {022102} (\bibinfo {year} {2011})}\BibitemShut {NoStop}%
\bibitem [{\citenamefont {Goldstein}(1980)}]{Goldstein}%
  \BibitemOpen
  \bibfield  {author} {\bibinfo {author} {\bibfnamefont {H.}~\bibnamefont
  {Goldstein}},\ }\href@noop {} {\emph {\bibinfo {title} {Classical
  Mechanics}}}\ (\bibinfo  {publisher} {Addison-Wesley},\ \bibinfo {year}
  {1980})\BibitemShut {NoStop}%
\bibitem [{\citenamefont {Chaté}\ and\ \citenamefont
  {Manneville}(1996)}]{Chate1996}%
  \BibitemOpen
  \bibfield  {author} {\bibinfo {author} {\bibfnamefont {H.}~\bibnamefont
  {Chaté}}\ and\ \bibinfo {author} {\bibfnamefont {P.}~\bibnamefont
  {Manneville}},\ }\bibfield  {title} {\bibinfo {title} {Phase diagram of the
  two-dimensional complex ginzburg-landau equation},\ }\href
  {https://doi.org/10.1016/0378-4371(95)00361-4} {\bibfield  {journal}
  {\bibinfo  {journal} {Physica A: Statistical Mechanics and its Applications}\
  }\textbf {\bibinfo {volume} {224}},\ \bibinfo {pages} {348–368} (\bibinfo
  {year} {1996})}\BibitemShut {NoStop}%
\bibitem [{\citenamefont {Ablowitz}\ and\ \citenamefont
  {Schober}(1994)}]{Ablowitz1994}%
  \BibitemOpen
  \bibfield  {author} {\bibinfo {author} {\bibfnamefont {M.}~\bibnamefont
  {Ablowitz}}\ and\ \bibinfo {author} {\bibfnamefont {C.}~\bibnamefont
  {Schober}},\ }\bibfield  {title} {\bibinfo {title} {{Effective chaos in the
  nonlinear Schr{\"o}dinger equation}},\ }\href@noop {} {\bibfield  {journal}
  {\bibinfo  {journal} {Contemporary Mathematics}\ }\textbf {\bibinfo {volume}
  {172}},\ \bibinfo {pages} {253} (\bibinfo {year} {1994})}\BibitemShut
  {NoStop}%
\bibitem [{\citenamefont {Frisch}(1995)}]{Frisch1995}%
  \BibitemOpen
  \bibfield  {author} {\bibinfo {author} {\bibfnamefont {U.}~\bibnamefont
  {Frisch}},\ }\href@noop {} {\emph {\bibinfo {title} {Turbulence: The Legacy
  of A.N. Kolmogorov}}}\ (\bibinfo  {publisher} {Cambridge University Press},\
  \bibinfo {year} {1995})\BibitemShut {NoStop}%
\bibitem [{\citenamefont {Tsekov}\ \emph {et~al.}(2017)\citenamefont {Tsekov},
  \citenamefont {Heifetz},\ and\ \citenamefont {Cohen}}]{Tsekov_2017}%
  \BibitemOpen
  \bibfield  {author} {\bibinfo {author} {\bibfnamefont {R.}~\bibnamefont
  {Tsekov}}, \bibinfo {author} {\bibfnamefont {E.}~\bibnamefont {Heifetz}},\
  and\ \bibinfo {author} {\bibfnamefont {E.}~\bibnamefont {Cohen}},\ }\bibfield
   {title} {\bibinfo {title} {Derivation of the local-mean stochastic quantum
  force},\ }\href@noop {} {\bibfield  {journal} {\bibinfo  {journal}
  {Fluctuations and Noise Letters}\ }\textbf {\bibinfo {volume} {16}},\
  \bibinfo {pages} {1750028} (\bibinfo {year} {2017})}\BibitemShut {NoStop}%
\bibitem [{\citenamefont {Heifetz}\ and\ \citenamefont
  {Plochotnikov}(2020)}]{Heifetz_2020}%
  \BibitemOpen
  \bibfield  {author} {\bibinfo {author} {\bibfnamefont {E.}~\bibnamefont
  {Heifetz}}\ and\ \bibinfo {author} {\bibfnamefont {I.}~\bibnamefont
  {Plochotnikov}},\ }\bibfield  {title} {\bibinfo {title} {Effective classical
  stochastic theory for quantum tunneling},\ }\href@noop {} {\bibfield
  {journal} {\bibinfo  {journal} {Phys. Lett. A}\ }\textbf {\bibinfo {volume}
  {384}},\ \bibinfo {pages} {126511} (\bibinfo {year} {2020})}\BibitemShut
  {NoStop}%
\bibitem [{\citenamefont {Nelson}(1966)}]{Nelson1966}%
  \BibitemOpen
  \bibfield  {author} {\bibinfo {author} {\bibfnamefont {E.}~\bibnamefont
  {Nelson}},\ }\bibfield  {title} {\bibinfo {title} {{Derivation of the
  Schr\"odinger equation from Newtonian mechanics}},\ }\href@noop {} {\bibfield
  {journal} {\bibinfo  {journal} {Phys. Rev.}\ }\textbf {\bibinfo {volume}
  {150}},\ \bibinfo {pages} {1079} (\bibinfo {year} {1966})}\BibitemShut
  {NoStop}%
\bibitem [{\citenamefont {Ramos}\ \emph {et~al.}(2020)\citenamefont {Ramos},
  \citenamefont {Spierings}, \citenamefont {Racicot},\ and\ \citenamefont
  {Steinberg}}]{Ramos_2020}%
  \BibitemOpen
  \bibfield  {author} {\bibinfo {author} {\bibfnamefont {R.}~\bibnamefont
  {Ramos}}, \bibinfo {author} {\bibfnamefont {D.}~\bibnamefont {Spierings}},
  \bibinfo {author} {\bibfnamefont {I.}~\bibnamefont {Racicot}},\ and\ \bibinfo
  {author} {\bibfnamefont {A.}~\bibnamefont {Steinberg}},\ }\bibfield  {title}
  {\bibinfo {title} {Measurement of the time spent by a tunneling atom within
  the barrier region},\ }\href@noop {} {\bibfield  {journal} {\bibinfo
  {journal} {Nature}\ }\textbf {\bibinfo {volume} {583}},\ \bibinfo {pages}
  {529} (\bibinfo {year} {2020})}\BibitemShut {NoStop}%
\bibitem [{\citenamefont {Aspect}(2015)}]{Aspect2015}%
  \BibitemOpen
  \bibfield  {author} {\bibinfo {author} {\bibfnamefont {A.}~\bibnamefont
  {Aspect}},\ }\bibfield  {title} {\bibinfo {title} {Closing the door on
  Einstein and Bohr's quantum debate},\ }\href@noop {} {\bibfield  {journal}
  {\bibinfo  {journal} {Physics}\ }\textbf {\bibinfo {volume} {8}},\ \bibinfo
  {pages} {123} (\bibinfo {year} {2015})}\BibitemShut {NoStop}%
\bibitem [{\citenamefont {Storz}\ \emph {et~al.}(2023)\citenamefont {Storz},
  \citenamefont {Schär}, \citenamefont {Kulikov} \emph {et~al.}}]{Storz2023}%
  \BibitemOpen
  \bibfield  {author} {\bibinfo {author} {\bibfnamefont {S.}~\bibnamefont
  {Storz}}, \bibinfo {author} {\bibfnamefont {J.}~\bibnamefont {Schär}},
  \bibinfo {author} {\bibfnamefont {A.}~\bibnamefont {Kulikov}}, \emph
  {et~al.},\ }\bibfield  {title} {\bibinfo {title} {Loophole-free Bell
  inequality violation with superconducting circuits},\ }\href@noop {}
  {\bibfield  {journal} {\bibinfo  {journal} {Nature}\ }\textbf {\bibinfo
  {volume} {617}},\ \bibinfo {pages} {265–270} (\bibinfo {year}
  {2023})}\BibitemShut {NoStop}%
\bibitem [{\citenamefont {'t~Hooft}(2017)}]{Hooft2017}%
  \BibitemOpen
  \bibfield  {author} {\bibinfo {author} {\bibfnamefont {G.}~\bibnamefont
  {'t~Hooft}},\ }\bibfield  {title} {\bibinfo {title} {Free will in the theory
  of everything},\ }\href@noop {} {\bibfield  {journal} {\bibinfo  {journal}
  {arXiv:1709.02874 [quant-ph]}\ } (\bibinfo {year} {2017})}\BibitemShut
  {NoStop}%
\bibitem [{\citenamefont {Hossenfelder}\ and\ \citenamefont
  {Palmer}(2020)}]{Hossenfelder2020}%
  \BibitemOpen
  \bibfield  {author} {\bibinfo {author} {\bibfnamefont {S.}~\bibnamefont
  {Hossenfelder}}\ and\ \bibinfo {author} {\bibfnamefont {T.}~\bibnamefont
  {Palmer}},\ }\bibfield  {title} {\bibinfo {title} {Rethinking
  superdeterminism},\ }\href@noop {} {\bibfield  {journal} {\bibinfo  {journal}
  {Front. Phys.}\ }\textbf {\bibinfo {volume} {8}},\ \bibinfo {pages} {139}
  (\bibinfo {year} {2020})}\BibitemShut {NoStop}%
\bibitem [{\citenamefont {Papatryfonos}\ \emph {et~al.}(2024)\citenamefont
  {Papatryfonos}, \citenamefont {Vervoort}, \citenamefont {Nachbin},
  \citenamefont {Labousse},\ and\ \citenamefont {Bush}}]{Bush2024}%
  \BibitemOpen
  \bibfield  {author} {\bibinfo {author} {\bibfnamefont {K.}~\bibnamefont
  {Papatryfonos}}, \bibinfo {author} {\bibfnamefont {L.}~\bibnamefont
  {Vervoort}}, \bibinfo {author} {\bibfnamefont {A.}~\bibnamefont {Nachbin}},
  \bibinfo {author} {\bibfnamefont {M.}~\bibnamefont {Labousse}},\ and\
  \bibinfo {author} {\bibfnamefont {J.~W.~M.}\ \bibnamefont {Bush}},\
  }\bibfield  {title} {\bibinfo {title} {Static Bell test in pilot-wave
  hydrodynamics},\ }\href {https://doi.org/10.1103/PhysRevFluids.9.084001}
  {\bibfield  {journal} {\bibinfo  {journal} {Phys. Rev. Fluids}\ }\textbf
  {\bibinfo {volume} {9}},\ \bibinfo {pages} {084001} (\bibinfo {year}
  {2024})}\BibitemShut {NoStop}%
\bibitem [{\citenamefont {Palmer}(2022)}]{Palmer2022}%
  \BibitemOpen
  \bibfield  {author} {\bibinfo {author} {\bibfnamefont {T.}~\bibnamefont
  {Palmer}},\ }\bibfield  {title} {\bibinfo {title} {Chaos theory eliminates
  quantum uncertainty},\ }\href@noop {} {\bibfield  {journal} {\bibinfo
  {journal} {IAI News}\ } (\bibinfo {year} {2022})},\ \bibinfo {note} {13
  October 2022}\BibitemShut {NoStop}%
\bibitem [{\citenamefont {Pismen}(1999)}]{Pismen1999}%
  \BibitemOpen
  \bibfield  {author} {\bibinfo {author} {\bibfnamefont {L.}~\bibnamefont
  {Pismen}},\ }\href@noop {} {\emph {\bibinfo {title} {Vortices in Nonlinear
  Fields}}}\ (\bibinfo  {publisher} {Oxford University Press},\ \bibinfo
  {address} {New York},\ \bibinfo {year} {1999})\BibitemShut {NoStop}%
\bibitem [{Note1()}]{Note1}%
  \BibitemOpen
  \bibinfo {note} {Away from the soliton's core, where $|\phi _s|\rightarrow
  0$, the phase of $\eta $ must be uniformly distributed over [$-\pi $,$\pi $]
  and, therefore, $\langle \theta (x,t)\theta (x',t')\rangle =\protect \frac
  {\pi ^2}{3}\delta (x-x')\delta (t-t')$.}\BibitemShut {Stop}%
\end{thebibliography}

%apsrev4-2.bst 2019-01-14 (MD) hand-edited version of apsrev4-1.bst
%Control: key (0)
%Control: author (8) initials jnrlst
%Control: editor formatted (1) identically to author
%Control: production of article title (0) allowed
%Control: page (0) single
%Control: year (1) truncated
%Control: production of eprint (0) enabled
%

\end{document}